\documentstyle[12pt]{article}
\newcommand\be{\begin{equation}}
\newcommand\ee{\end{equation}}
\newcommand\bea{\begin{eqnarray}}
\newcommand\eea{\end{eqnarray}}
\newcommand\ket[1]{|#1\rangle}

\newcommand{\fatalpha}{{\bf \alpha \kern -0.44em \alpha}}
\newcommand{\fatsigma}{{\bf \sigma \kern -0.54em \sigma}}
\newcommand{\tpchi}{{\bf \chi \kern -0.35em \chi}}
\newcommand{\llambda}{{\bf \lambda \kern -0.45em \lambda}}

\bibliography{plain}
\title{\bf Entanglement entropy in the Hamming networks}\vspace{20mm}
\author{ M. A. Jafarizadeh$^{a}$
 \thanks{E-mail:jafarizadeh@tabrizu.ac.ir}
  S. Nami$^{a}$
 \thanks{E-mail:S.Nami@tabrizu.ac.ir}
 F. Eghbalifam$^{a}$
 \thanks{E-mail:F.Egbali@tabrizu.ac.ir},
\\ $^a${\small Department of Theoretical Physics and Astrophysics,
University of Tabriz, Tabriz 51664, Iran.}} \pagebreak

\vspace{20mm}
\begin{document}
\maketitle \vspace{15mm}

\begin{abstract}
We investigate the Hamming networks that their nodes are
considered as quantum harmonic oscillators. The entanglement of
the ground state can be used to quantify the amount of information
each part of a network shares with the rest of the system via
quantum fluctuations. Therefore, the Schmidt numbers and
entanglement entropy between two special parts of Hamming network,
can be calculated. To this aim, first we use the stratification
method to rewrite the adjacency matrix of the network in the
stratification basis. Then the entanglement entropy and Schmidt
number for special partitions are calculated analytically by using
the generalized Schur complement method. Also, we calculate the
entanglement entropy between two arbitrary subsets (two equal
subsets have the same number of vertices) in $H(2,3)$  and
$H(2,4)$ numerically, and we give the minimum and maximum values
of entanglement entropy in these two Hamming network.
\end{abstract}

\newpage
\section{Introduction}
Entanglement plays a crucial role in quantum information
processing, including quantum communication [1,2] and quantum
computation [3-5]. It is one of the remarkable features that
distinguishes quantum mechanics from classical mechanics.\\ For
decades, entanglement has been the focus of much work in the
foundations of quantum mechanics, being associated particularly
with quantum nonseparability and the violation of Bells
inequalities [6]. Since entanglement has become regarded as such
an important resource, there is a need for a means of quantifying
it. For the case of bipartite entanglement, a recent exhaustive
review was written by the Horodecki family [7] and entanglement
measures have been reviewed in detail by Virmani and Plenio [8].
One of the operational entanglement criteria is the Schmidt
decomposition [9-11]. The Schmidt decomposition is a very good
tool to study entanglement of bipartite pure states. The Schmidt
number provides an important variable to classify entanglement.
The entanglement of a partly entangled pure state can be naturally
parametrized by its entropy of entanglement, defined as the von
Neumann entropy, or equivalently as the Shannon entropy of the
squares of the Schmidt coefficients [9-11]. The situation
simplifies if only so called \textit{Gaussian states} of the
harmonic oscillator modes are considered [12-17]. The importance
of gaussian states is two-fold; firstly, its structural
mathematical description makes them much more amenable than any
other continuous variable system (Continuous variable systems are
those described by canonical conjugated coordinates $x$  and $p$
endowed with infinite dimensional Hilbert spaces). Secondly, its
production, manipulation and detection with current optical
technology can be done with a very high degree of accuracy and
control. In [18] the authors quantified the amount of information
that a single element of a quantum network shares with the rest of
the system. They considered a network of quantum harmonic
oscillators and analyzed its ground state to compute the entropy
of entanglement that vacuum fluctuations creates between single
nodes and the rest of the network by using the Von Neumann
entropy. In [19], jafarizadeh et al, quantify the entanglement
entropy between two parts of the network. To this aim,they compute
the vacuum state of bosonic modes harmonically coupled through the
specific adjacency matrix of a given network. \\In this paper, we
want can calculate the Schmidt numbers and entanglement entropy
between two special parts of Hamming network by generalized Schur
complement method.
\\In  section II, we give some
preliminaries such as definitions related to association schemes,
corresponding stratification and The Terwilliger algebra.
\\In section III, the generalized Schur complement method is used for calculating the Schmidt numbers and entanglement entropy.
In this method, we will apply the generalized Schur complement
method to the potential matrix in the stratification basis several
times to calculate the entanglement entropy between two arbitrary
parts in Hamming network. Then ,by using this method, the
entanglement entropy and Schmidt number for special partitions are
calculated analytically. Also, we calculate the entanglement
entropy between two arbitrary subsets (two equal subsets have the
same number of vertices) in $H(2,3)$  and $H(2,4)$ numerically,
and we give the minimum and maximum values of entanglement entropy
in these two Hamming network.
\section{Preliminaries} In this section we give some
preliminaries such as definitions related to association schemes,
corresponding stratification and The Terwilliger algebra.

\subsection{The model and Hamiltonian}
We consider nodes as identical quantum oscillators, interacting as
dictated by the network topology encoded in the Laplacian $L$. The
Laplacian of a network is defined from the Adjacency matrix as
$L_{ij} = k_i\delta_{ij}- A_{ij}$ , where $k_i =\sum_j A_{ij}$ is
the connectivity of node $i$, i.e., the number of nodes connected
to $i$. The Hamiltonian of the quantum network thus reads:
\begin{equation}
H=\frac{1}{2}(P^T P+ X^T(I+2gL)X)
\end{equation}
here $I$ is the $N \times N$ identity matrix, $g$ is the coupling
strength between connected oscillators while $p^T=(p_1,p_2,...,
p_N)$ and $x^T=(x_1,x_2, ..., x_N)$ are the operators
corresponding to the momenta and positions of nodes respectively,
satisfying the usual commutation relations: $[x, p^T] = i\hbar I$
(we set $\hbar = 1$ in the following) and the matrix $V=I+2gL$ is
the potential matrix. Then the ground state of this Hamiltonian
is:
\begin{equation}
\psi(X)=\frac{(det(I+2gL))^{1/4}}{\pi^{N/4}}exp(-\frac{1}{2}(X^T(I+2gL)X))
\end{equation}
Where the $A_g=\frac{(det(I+2gL))^{1/4}}{\pi^{N/4}}$ is the
normalization factor for wave function. The elements of the
potential matrix in terms of entries of adjacency matrix is
$$V_{ij}=(1+2g\kappa_i)\delta_{ij}-2gA_{ij}$$

\subsection{Schmidt decomposition and entanglement entropy}

Any bipartite pure state $|\psi\rangle_{AB} \in
\textsl{H}=\textsl{H}_A \otimes\textsl{H}_B$ can be decomposed, by
choosing an appropriate basis, as
\begin{equation}
|\psi\rangle_{AB}=\sum_{i=1}^m
\alpha_i|a_i\rangle\otimes|b_i\rangle
\end{equation}
where $1 \leq m \leq min\{dim(\textsl{H}_A); dim(\textsl{H}_B)\}$,
and $\alpha_i
> 0$ with $\sum_{i=1}^m \alpha_i^2 = 1$. Here $|a_i\rangle$ ($|b_i\rangle$) form a part of an
orthonormal basis in $\textsl{H}_A$ ($\textsl{H}_B$). The positive
numbers $\alpha_i$ are called the Schmidt coefficients of
$|\psi\rangle_{AB}$ and the number $m$ is called the Schmidt rank
of $|\psi\rangle_{AB}$.

Entropy of entanglement is defined as the von Neumann entropy of
either $\rho_A$ or $\rho_B$:
\begin{equation}
E=-Tr\rho_A log_2\rho_A= Tr\rho_B log_2\rho_B=-\sum_i\alpha_i^2
log_2 \alpha_i^2
\end{equation}

\subsection{Association scheme}
First we recall the definition of association schemes. The reader
is referred to Ref.[?], for further information on association
schemes.

\textbf{Definition 2.1} (Symmetric association schemes). Let $V$
be a set of vertices, and let $R_i(i = 0, 1, ..., d)$ be nonempty
relations on $V$ (i.e., subset of $V\times V$ ). Let the following
conditions (1), (2), (3) and (4) be satisfied. Then, the relations
$\{R_i\}_{0\leq i\leq d}$ on $V\times V$ satisfying the following
conditions

(1) $\{R_i\}_{0\leq i\leq d}$ is a partition of $V\times V$

(2) $R_0 = {(\alpha,\alpha) : \alpha \in V }$

(3) $R_i = R^t_i$ for $0\leq i \leq d$, where
$R^t_i={(\beta,\alpha) : (\alpha,\beta)\in R_i}$

(4) For $(\alpha,\beta)\in R_k$, the number $p^k_{ij}=|{\gamma \in
V : (\alpha,\gamma)\in R_i \quad and \quad (\gamma,\beta)\in
R_j}|$ does not depend on $(\alpha,\beta)$ but only on $i,j$ and
$k$, define a symmetric association scheme of class $d$ on $V$
which is denoted by $Y = (V, \{R_i\}_{0\leq i\leq d})$.
Furthermore, if we have $p^k_{ij} = p^k_{ji}$ for all $i, j, k =
0, 2, ..., d$, then $Y$ is called commutative.

The number $v$ of the vertices, $|V|$, is called the order of the
association scheme and $R_i$ is called $i$-th relation.

The intersection number $p^k_{ij}$ can be interpreted as the
number of vertices which have relation $i$ and $j$ with vertices
$\alpha$ and $\beta$, respectively provided that $(\alpha,\beta)
\in R_k$, and it is the same for all elements of relation $R_k$.
For all integers $i (0\leq i\leq d)$, set $\kappa_i = p^0_{ii}$
and note that $\kappa_i \neq 0$, since $R_i$ is non-empty. We
refer to $\kappa_i$ as the $i$-th valency of $Y$.

Let $Y = (X, \{R_i\}_{0\leq i\leq d})$ be a commutative symmetric
association scheme of class $d$, then the matrices $A_0,A_1,
...,A_d$ defined by
\begin{equation}\label{adj.}
    \bigl(A_{i})_{\alpha, \beta}\;=\left\{\begin{array}{c}
      \hspace{-2.3cm}1 \quad \mathrm{if} \;(\alpha,
    \beta)\in R_i, \\
      0 \quad \mathrm{otherwise} \quad \quad \quad(\alpha, \beta
    \in V) \\
    \end{array}\right.
\end{equation}
are adjacency matrices of $Y$ such that
\begin{equation}\label{ss}
A_iA_j=\sum_{k=0}^{d}p_{ij}^kA_{k}.
\end{equation}
From (\ref{ss}), it is seen that the adjacency matrices $A_0, A_1,
..., A_d$ form a basis for a commutative algebra \textsf{A} known
as the Bose-Mesner algebra of $Y$. This algebra has a second basis
$E_0,..., E_d$ (known as primitive idempotents of $Y$) so that
\begin{equation}\label{idem}
E_0 = \frac{1}{n}J, \;\;\;\;\;\;\ E_iE_j=\delta_{ij}E_i,
\;\;\;\;\;\;\ \sum_{i=0}^d E_i=I.
\end{equation}
where, $J$ is the all-one matrix in $\textsf{A}$. Let $P$ and $Q$
be the matrices relating the two bases for $\textsf{A}$:
$$
A_i=\sum_{i=0}^d P_{ij}E_j, \;\;\;\;\ 0\leq j\leq d,
$$
\begin{equation}\label{m2}
E_i=\frac{1}{n}\sum_{i=0}^d Q_{ij}A_j, \;\;\;\;\ 0\leq j\leq d.
\end{equation}
Then clearly
\begin{equation}\label{pq}
PQ=QP=nI.
\end{equation}
It also follows that
\begin{equation}\label{eign}
A_iE_j=P_{ij}E_j,
\end{equation}
which shows that the $P_{ij}$ (resp. $Q_{ij}$) is the $j$-th
eigenvalue (resp. the $j$-th dual eigenvalue ) of $A_i$ (resp.
$E_i$) and that the columns of $E_j$ are the corresponding
eigenvectors. Thus, $m_i=$rank$(E_i)$ is the multiplicity of the
eigenvalue $P_{ij}$ of $A_i$ (provided that $P_{ij}\neq P_{kj}$
for $k \neq i$). We see that $m_0=1, \sum_i m_i=n$, and
$m_i=$trace$E_i=n(E_i)_{jj}$ (indeed, $E_i$ has only eigenvalues
$0$ and $1$, so rank($E_k$) equals the sum of the eigenvalues).

Clearly, each non-diagonal (symmetric) relation $R_i$ of an
association scheme $Y=(V,\{R_i\}_{0\leq i\leq d})$ can be thought
of as the network $(V,R_i)$ on $V$, where we will call it the
underlying network of association scheme $Y$. In other words, the
underlying network $\Gamma=(V,R_1)$ of an association scheme is an
undirected connected network, where the set $V$ and $R_1$ consist
of its vertices and edges, respectively. Obviously replacing $R_1$
with one of the other relations such as $R_i$, for  $i\neq 0,1$
will also give us an underlying network $\Gamma=(V,R_i)$ (not
necessarily a connected network) with the same set of vertices but
a new set of edges $R_i$.

\subsection{Stratification} For an underlying network $\Gamma$, let
$W={\mathcal{C}}^n$ (with $n=|V|$) be the vector space over
$\mathcal{C}$ consisting of column vectors whose coordinates are
indexed by vertex set $V$ of $\Gamma$, and whose entries are in
$\mathcal{C}$. For all $\beta\in V$, let $\ket{\beta}$ denotes the
element of $W$ with a $1$ in the $\beta$ coordinate and $0$ in all
other coordinates. We observe $\{\ket{\beta} | \beta\in V\}$ is an
orthonormal basis for $W$, but in this basis, $W$ is reducible and
can be reduced to irreducible subspaces $W_i$, $i=0,1,...,d$,
i.e.,
\begin{equation}
W=W_0\oplus W_1\oplus...\oplus W_d,
\end{equation}
where, $d$ is diameter of the corresponding association scheme. If
we define
 $\Gamma_i(o)=\{\beta\in V:
(o, \beta)\in R_i\}$ for an arbitrary chosen vertex $o\in V$
(called reference vertex), then, the vertex set $V$ can be written
as disjoint union of $\Gamma_i(\alpha)$, i.e.,
 \begin{equation}\label{asso1}
 V=\bigcup_{i=0}^{d}\Gamma_{i}(\alpha).
 \end{equation}
In fact, the relation (\ref{asso1}) stratifies the network into a
disjoint union of strata (associate classes) $\Gamma_{i}(o)$. With
each stratum $\Gamma_{i}(o)$ one can associate a unit vector
$\ket{\phi_{i}}$ in $W$ (called unit vector of $i$-th stratum)
defined by
\begin{equation}\label{unitv}
\ket{\phi_{i}}=\frac{1}{\sqrt{\kappa_{i}}}\sum_{\alpha\in
\Gamma_{i}(o)}\ket{\alpha},
\end{equation}
where, $\ket{\alpha}$ denotes the eigenket of $\alpha$-th vertex
at the associate class $\Gamma_{i}(o)$ and
$\kappa_i=|\Gamma_{i}(o)|$ is called the $i$-th valency of the
network ($\kappa_i:=p^0_{ii}=|\{\gamma:(o,\gamma)\in
R_i\}|=|\Gamma_{i}(o)|$). For $0\leq i\leq d$, the unit vectors
$\ket{\phi_{i}}$ of Eq.(\ref{unitv}) form a basis for irreducible
submodule of $W$ with maximal dimension denoted by $W_0$. Since
$\{\ket{\phi_{i}}\}_{i=0}^d$ becomes a complete orthonormal basis
of $W_0$, we often write
\begin{equation}
W_0=\sum_{i=0}^d\oplus \textbf{C}\ket{\phi_{i}}.
\end{equation}

Let $A_i$ be the adjacency matrix of the underlying network
$\Gamma$. From the action of $A_i$ on reference state
$\ket{\phi_0}$ ($\ket{\phi_0}=\ket{o}$, with $o\in V$ as reference
vertex), we have
\begin{equation}\label{Foc1}
A_i\ket{\phi_0}=\sum_{\beta\in \Gamma_{i}(o)}\ket{\beta}.
\end{equation}
 Then by using (\ref{unitv}) and (\ref{Foc1}),
 we obtain
\begin{equation}\label{Foc2}
A_i\ket{\phi_0}=\sqrt{\kappa_i}\ket{\phi_i}.
\end{equation}

\section{Entanglement entropy between two parts of a network}
In order to calculate the entanglement entropy between two equal
parts in the graph (half first strata are in one subset and the
other strata are in the second subset), we introduce the following
process:

First we want to generalize Schur complement method [19]: Suppose
we have the following matrix which is composed of block matrices.
\begin{equation}
V=\left(\begin{array}{ccc}
          V_{11}& V_{12} & 0\\
            V_{21}& V_{22} & V_{23}\\
            0 & V_{32} & V_{33}\\
          \end{array}\right)
\end{equation}
Then we do the generalized Schur complement transformation. This
transformation is:
\begin{equation}
\left(\begin{array}{ccc}
          V_{11}& V_{12} & 0\\
            V_{21}& V_{22} & V_{23}\\
            0 & V_{32} & V_{33}\\
          \end{array}\right)=\left(\begin{array}{ccc}
          1& 0 & 0\\
            0& 1& V_{23}V_{33}^{-1}\\
            0 & 0 & 1\\
          \end{array}\right)\left(\begin{array}{ccc}
          V_{11}& V_{12} & 0\\
            V_{12}^T& V_{22}-V_{23}V_{33}^{-1}V_{32} & 0\\
            0 & 0 & V_{33}\\
          \end{array}\right)\left(\begin{array}{ccc}
          1& 0 & 0\\
            0& 1& 0\\
            0 & V_{33}^{-1}V_{32} & 1\\
          \end{array}\right)
\end{equation}
In our work, we will apply the generalized Schur complement method
to the potential matrix in the stratification basis several times.
So in the transformed matrix all of the blocks are scalar.
Therefore the potential matrix is transformed to a $2\times 2$
matrix finally.
\begin{equation}
V=\left(\begin{array}{cc}
         a_{11}&a_{12} \\
          a^T_{12}& a_{22}\\
          \end{array}\right)
\end{equation}
The wave function in this stage is
\begin{equation}
\psi(x,y)=A_g exp(-\frac{1}{2}(x\quad \quad
y)\left(\begin{array}{cc}
          a_{11}&a_{12}\\
            a_{12}& a_{22}\\
          \end{array}\right)\left(\begin{array}{c}
          x\\
            y\\
          \end{array}\right))
\end{equation}
by rescaling the variables $x$ and $y$:
$$\widetilde{x}=a_{11}^{1/2}x$$
$$\widetilde{y}=a_{22}^{1/2}y$$
the ground state wave function is transformed to
\begin{equation}
\psi(\widetilde{x},\widetilde{y})=A_g
exp(-\frac{1}{2}(\widetilde{x}\quad \quad
\widetilde{y})\left(\begin{array}{cc}
          1& \gamma\\
            \gamma & 1\\
          \end{array}\right)\left(\begin{array}{c}
          \widetilde{x}\\
            \widetilde{y}\\
          \end{array}\right))
\end{equation}
where
\begin{equation}
\gamma=a_{11}^{-1/2}a_{12}a_{22}^{-1/2}
\end{equation}
So the ground state wave function is
\begin{equation}
\psi(\widetilde{x},\widetilde{y})=A_g
e^{-\frac{\widetilde{x}^2}{2}-\frac{\widetilde{y}^2}{2}-\gamma
\widetilde{x}\widetilde{y}}
\end{equation}

Then we can use following identity to calculate the schmidt number
of this wave function,
\begin{equation}
\frac{1}{\pi^{1/2}}exp(-\frac{1+t^2}{2(1-t^2)}((x)^2+(y)^2))+\frac{2t}{1-t^2}x
y)=(1-t^2)^{1/2}\sum_n t^n \psi_n(x)\psi_n(y)
\end{equation}
In order to calculating the entropy, we apply a change of variable
as
$$1-t^2=\frac{2}{\nu+1}$$
$$t^2=\frac{\nu-1}{\nu+1}$$
So the above identity becomes
\begin{equation}
\frac{1}{\pi^{1/2}}exp(-\frac{\nu}{2}((x)^2+(y)^2))+(\nu^2-1)^{1/2}x
y)=(\frac{2}{\nu+1})^{1/2}\sum_n (\frac{\nu-1}{\nu+1})^{n/2}
\psi_n(x)\psi_n(y)
\end{equation}
and the reduced density matrix is
\begin{equation}
\rho=\frac{2}{\nu+1}\sum_{n}(\frac{\nu-1}{\nu+1})^n |n\rangle
\langle n|
\end{equation}
the entropy is
\begin{equation}
S(\rho)=-\sum_n p_n log(p_n)
\end{equation}
where $p_n=\frac{2}{\nu+1}(\frac{\nu-1}{\nu+1})^n$
\begin{equation}
\sum_n p_n log(p_n)=log(\frac{2}{\nu+1})+\langle n \rangle
log(\frac{\nu-1}{\nu+1})
\end{equation}
and $\langle n \rangle = \frac{\nu -1}{2}$

\begin{equation}
S(\rho)=\frac{\nu +1}{2} log(\frac{\nu +1}{2})-\frac{\nu
-1}{2}log(\frac{\nu -1}{2})
\end{equation}
By comparing the wave function (3-24) and the identity (3-25) and
define the scale $\mu^2$, we conclude that
$$\nu=1 \times \mu^2$$
$$(\nu^2-1)^{1/2}=-\gamma \times \mu^2$$
After some straightforward calculation we obtain
\begin{equation}
\nu=(\frac{1}{1-\gamma^2})^{1/2}
\end{equation}
By above discussion we conclude that
$$e^{-\frac{(x)^2}{2}-\frac{(y)^2}{2}-\gamma xy}=\sum_n \lambda_{n}\psi_n(x)\psi_n(y)$$
where
$\lambda_{n}=(\frac{2}{\nu+1})^{1/2}(\frac{\nu-1}{\nu+1})^{n/2}$.

and the entropy is
\begin{equation}
S(\rho)=\frac{\nu +1}{2} log(\frac{\nu +1}{2})-\frac{\nu
-1}{2}log(\frac{\nu -1}{2})
\end{equation}
\subsection{Hamming network}
The symmetric product of $d$-tuples of trivial scheme $K_n$ with
adjacency matrices of $I_n$, $J_n - I_n$ is association scheme
with the following adjacency matrices ( generators of its
Bose-Mesner algebra)

$$A_{0}=I_{n}\otimes I_{n}\otimes ... \otimes I_{n}$$
$$A_{1}=\sum_{permutation}(J_{n}-I_{n})\otimes I_{n}\otimes...\otimes
I_{n}$$
$$\vdots$$
\begin{equation}
A_{i}=\sum_{permutation}\underbrace{(J_{n}-I_{n})\otimes(J_{n}-I_{n})\otimes
...\otimes(J_{n}-I_{n})}_{i}\otimes I_{n}\otimes...\otimes I_{n}
\end{equation}

Where $J_{n}$ is $n\times n$ matrix with all matrix elements equal
to one. This scheme is the well known Hamming scheme with
intersection number
$$a_i=\frac{(n-1)^id(d-1)\ldots(d-i+1)}{i!},\quad\quad\quad 1\leq i\leq d$$
$$b_i=i,\quad\quad\quad 1\leq i\leq d,$$
\begin{equation}
c_i=(n-1)(d-i), \quad\quad\quad  0\leq i\leq d-1,
\end{equation}

where its underlying graph is the cartesian product of $d$-tuples
of cyclic group $Z_n$. We can rewrite the matrix $J_{n}-I_{n}$ as
the following and then calculate the singular value decomposition
of matrix $B$

$$
\left(\begin{array}{cc}
          0& B\\
           B^T & J_{n-1}-I_{n-1}\\
          \end{array}\right)\equiv \left(\begin{array}{cc}
          0& \begin{array}{cccc}1&1& \ldots &1\\ \end{array}\\
           \begin{array}{c}1\\1\\ \vdots \\1\\ \end{array} & J_{n-1}-I_{n-1}\\
          \end{array}\right)=
$$
\begin{equation}
=\left(\begin{array}{cc} 1&0\\
0&O_{n-1}\\
\end{array}\right)
 \left(\begin{array}{cc}
          0& \begin{array}{cccc}\sqrt{n-1}&0& \ldots &0\\ \end{array}\\
           \begin{array}{c}\sqrt{n-1}\\0\\ \vdots \\0\\ \end{array} & \begin{array}{cccc}n-2&0&\ldots &0\\
           0&-1&\ldots &0\\
           \vdots &\vdots &\ddots& \vdots \\
           0&0& \ldots &-1\\ \end{array}\\
          \end{array}\right)\left(\begin{array}{cc} 1&0\\
0&O_{n-1}\\
\end{array}\right)
\end{equation}

Now we want to introduce new basis proportional to this
transformation
$$|\widetilde{0}\rangle=|0\rangle$$
$$|\widetilde{1}\rangle=\frac{1}{\sqrt{n-1}}(|1\rangle+|2\rangle+\ldots+|n-1\rangle)$$
\begin{equation}
|\widetilde{k}\rangle=\frac{1}{\sqrt{n-1}}(\sum_{j=1}^{n-1}w^{(k-1)(j-1)}|j\rangle)\quad\quad
k=2,3,\ldots,n-1
\end{equation}

The effect of matrix $J_n-I_n$ on the above basis from equation
(3-35) will be
$$(J_n-I_n)|\widetilde{0}\rangle=\sqrt{n-1}|\widetilde{1}\rangle$$
$$(J_n-I_n)|\widetilde{1}\rangle=(n-2)|\widetilde{1}\rangle+\sqrt{n-1}|\widetilde{0}\rangle$$
\begin{equation}
(J_n-I_n)|\widetilde{k}\rangle=-|\widetilde{k}\rangle,\quad\quad\quad
k=2,3,\ldots, n-1
\end{equation}

So the matrices $(\widetilde{J_n-I_n})_{-}$,
$(\widetilde{J_n-I_n})_{0}$ and $(\widetilde{J_n-I_n})_{+}$ are
$$(\widetilde{J_n-I_n})_{-}=\sqrt{n-1}|\widetilde{1}\rangle \langle \widetilde{0}|$$
$$(\widetilde{J_n-I_n})_{0}=(n-2)|\widetilde{1}\rangle \langle \widetilde{1}|-\sum_{k=2}^{n-1}|\widetilde{k}\rangle \langle \widetilde{k}|$$
\begin{equation}
(\widetilde{J_n-I_n})_{+}=\sqrt{n-1}|\widetilde{0}\rangle
\langle\widetilde{1}|
\end{equation}

Then we introduce the matrices $A_{+}$, $A_{-}$ and $A_0$ as
$$A_{+}=\sum I_n\otimes \ldots \otimes I_n \otimes (\widetilde{J_n-I_n})_{+} \otimes I_n \otimes \ldots \otimes I_n$$
$$A_{-}=\sum I_n\otimes \ldots \otimes I_n \otimes (\widetilde{J_n-I_n})_{-} \otimes I_n \otimes \ldots \otimes I_n$$
\begin{equation}
A_{0}=\sum I_n\otimes \ldots \otimes I_n \otimes
(\widetilde{J_n-I_n})_{0} \otimes I_n \otimes \ldots \otimes I_n
\end{equation}

In this stage we want to construct the first stratum of Hamming
network. We begin with
$$|\phi_0\rangle_{1}=|\widetilde{00\ldots0}\rangle=|\widetilde{0}\rangle \otimes |\widetilde{0}\rangle \otimes \ldots \otimes
|\widetilde{0}\rangle$$

By using the equation (3-38), we conclude that
$$A_{+}|\phi_0\rangle_{1}=0$$
$$A_{-}|\phi_0\rangle_{1}=\sqrt{(n-1)d}|\phi_1\rangle_{1}$$
So we have
$$|\phi_1\rangle_{1}=\frac{1}{\sqrt{d}}(|\widetilde{1}\widetilde{0}\ldots \widetilde{0}\rangle+|\widetilde{0}\widetilde{1}\ldots \widetilde{0}\rangle+\ldots+|\widetilde{0}\widetilde{0}\ldots \widetilde{1}\rangle)=\frac{1}{\sqrt{d}}\sum_i|\widetilde{0}\ldots \widetilde{0}\underbrace{\widetilde{1}}_i\widetilde{0}\ldots \widetilde{0}\rangle$$
$$|\phi_2\rangle_{1}=\frac{1}{\sqrt{C^d_2}}\sum_{i,j=1}^d|\widetilde{0}\ldots \widetilde{0}\underbrace{\widetilde{1}}_i\widetilde{0}\ldots \widetilde{0}\underbrace{\widetilde{1}}_j\widetilde{0}\ldots \widetilde{0}\rangle$$
\begin{equation}
|\phi_j\rangle_{1}=\frac{1}{\sqrt{C^d_j}}\sum_{i_1<i_2<\ldots<i_d=1}^d|\widetilde{0}\ldots
\widetilde{0}\underbrace{\widetilde{1}}_{i_1}\widetilde{0}\ldots
\widetilde{0}\underbrace{\widetilde{1}}_{i_2}\widetilde{0}\ldots
\widetilde{0}\underbrace{\widetilde{1}}_{i_d}\widetilde{0}\ldots
\widetilde{0}\rangle
\end{equation}
We are interesting in the effect of $A_{-}$ on the one stratum of
first strata $|\phi_j\rangle_{1}$,
\begin{equation}
A_{-}|\phi_j\rangle_{1}=\sqrt{n-1}\sqrt{(j+1)(d-j)}|\phi_{j+1}\rangle_{1}
\end{equation}
from the above equation we conclude that, without the coefficient
$\sqrt{n-1}$, this graph is the spin $1/2$ representation of the
angular momentum operator.

Also we have
$$A_{0}|\phi_j\rangle_{1}=j(n-2)|\phi_{j}\rangle_{1}$$
\begin{equation}
A_{+}|\phi_j\rangle_{1}=\sqrt{n-1}\sqrt{(j+1)(d-j)}|\phi_{j-1}\rangle_{1}
\end{equation}

The above basis have only the elements $0$, $1$. For
$\alpha_{i},i=2,3,...,n-1$. We know that, the number of all states
are $n^{d}$. If all elements are consisted from $0,1$, we have,
$2^{d}$ states. Now, imagine one of the elements of state become
$\alpha_{i}$. So, $\alpha_{i}$ can take $n-2$ value and can be in
$C^{d}_{1}$ place, therefore the number of states with one
parameter $\alpha_{i}$, is $(n-2)C^{d}_{1}$. Therefore,we have
$2^{d-1}$ states of this kind and each state, is repeated by
$(n-2)C^{d}_{1}$. So, the number of all states with one
$\alpha_{i}$ is $2^{d-1}(n-2)C^{d}_{1}$. By the same way, let we
have two different $\alpha_i$ and $\alpha_j$, then the number of
states will be $2^{d-2}(n-2)^{2}C^{d}_{2}$. Similarity, if we have
$d$ different $\alpha_{i}$,then we have $(n-2)^{d}$ states, that
all of them are singlet. We can extend the term $n^d$, as
following
\begin{equation}
n^{d}=2^{d}+2^{d-1}(n-2)C^{d}_{1}+2^{d-2}(n-2)^{2}C^{d}_{2}+...+2^{d-m}(n-2)^{m}C^{d}_{m}+...+2(n-2)^{d-1}C^{d}_{d-1}+(n-2)^{d}\end{equation}
The first term in the above is the number of all states contained
only $0$ and $1$, i.e. there is not the parameter $\alpha_{i}$ in
them. The second term in the extension is the number of all
states which have only one $\alpha_{i}$. In fact the $m$th term
in the above, is the number of all states that these states are
contained $m$ different $\alpha_{i}$s.

The adjacency matrix for these states with $m$ different
$\alpha_i$ is similar to the adjacency matrix for the
$|\phi_i\rangle$s, but it will be for $d'=d-m$, i.e., the effect
of operators $A_{+}$ and $A_{-}$ on the states
($|\varphi_j\rangle$) is similar to the binary states
($|\phi_j\rangle$). The diagonal elements can be extract from
following equation:
\begin{equation}
A_0|\varphi_j\rangle_{\alpha_1,\alpha_2,\ldots,\alpha_m}=(j(n-2)-m)|\varphi_j\rangle_{\alpha_1,\alpha_2,\ldots,\alpha_m}\quad\quad
j=0,1,\ldots,d-m
\end{equation}

For other strata, we begin with
$|\phi_1\rangle_{k}=\frac{1}{\sqrt{d}}\sum (w_d)^{kj}|0\ldots
0\underbrace{1}_j0\ldots 0\rangle$, then apply the operator
$A_{-}$ to this state and obtain the state $|\phi_2\rangle_{k}$,
So
$$|\phi_1\rangle_{k}=\frac{1}{\sqrt{d}}\sum (w_d)^{kj}|0\ldots
0\underbrace{1}_j0\ldots 0\rangle$$
$$|\phi_2\rangle_{k}=\frac{1}{\sqrt{d(d-2)}}\sum_{i_1< i_2}^d(w^{ki_1}+w^{ki_2})|0\ldots 0\underbrace{1}_{i_1}0\ldots0\underbrace{1}_{i_2}0\ldots0\rangle$$
$$|\phi_3\rangle_{k}=\frac{1}{\sqrt{d(d-2)(d-3)/2}}\sum_{i_1< i_2< i_3}^d(w^{ki_1}+w^{ki_2}+w^{ki_3})|0\ldots 0\underbrace{1}_{i_1}0\ldots0\underbrace{1}_{i_2}0\ldots0\underbrace{1}_{i_3}0\ldots0\rangle$$
$$\vdots$$
\begin{equation}
|\phi_m\rangle_{k}=\frac{1}{\sqrt{\frac{d(d-2)\ldots(d-m)}{(m-1)!}}}\sum_{i_1<
i_2<\ldots< i_m}^d(w^{ki_1}+w^{ki_2}+\ldots+w^{ki_m})|0\ldots
0\underbrace{1}_{i_1}0\ldots0\underbrace{1}_{i_2}0\ldots0\underbrace{1}_{i_m}0\ldots0\rangle
\end{equation}
It can be shown that for these states
\begin{equation}
A_{-}|\phi_m\rangle_{k}=\sqrt{n-1}\sqrt{m(d-m-1)}|\phi_{m+1}\rangle_{k}
\end{equation}
These states are equivalent to the angular momentum $2J=d-2$. For
the third stratum we have
\begin{equation}
|\phi_1\rangle_{k_1,k_2}=\frac{1}{2d}\sum_{i_1\neq
i_2}^d(w^{k_1i_1+k_2i_2}-w^{k_1i_2+k_2i_1})|0\ldots
0\underbrace{1}_{i_1}0\ldots0\underbrace{1}_{i_2}0\ldots0\rangle
\end{equation}
These states are equivalent to the representation of $2J=d-4$
angular momentum. The highest weight of these kind of basis, is
$$|\phi_1\rangle_{k_1,k_2,\ldots,k_m}=$$
\begin{equation}
\frac{1}{m!\sqrt{d^m}}\sum_{i_1<i_2<\ldots<i_m=1}^d
\sum_{\alpha_1,\alpha_2,\ldots,\alpha_m}(w^{k_{\alpha_1}i_1+k_{\alpha_2}i_2+\ldots+k_{\alpha_m}i_m})|0\ldots0\underbrace{1}_{i_1}0\ldots0\underbrace{1}_{i_2}0\ldots0\underbrace{1}_{i_m}0\ldots0\rangle
\end{equation}
For the network with an arbitrary $d$, the number of strata is
$[\frac{d}{2}]+1$.

For calculating the entanglement entropy, we use the following
equations
$$A_{-}|\phi_j\rangle_{1}=\sqrt{n-1}\sqrt{(j+1)(d-j)}|\phi_{j+1}\rangle_{1}$$
$$A_{0}|\phi_j\rangle_{1}=j(n-2)|\phi_{j}\rangle_{1}$$
\begin{equation}
A_{+}|\phi_j\rangle_{1}=\sqrt{n-1}\sqrt{(j+1)(d-j)}|\phi_{j-1}\rangle_{1}
\end{equation}

So the Adjacency matrix of each strata is
\begin{equation}
A=\left(\begin{array}{cccccc}
         0& c\sqrt{d'} & 0&0&\ldots &0\\
            c\sqrt{d'} & n-2 & c\sqrt{2(d'-1)}&0&\ldots &0\\
            0 & c\sqrt{2(d'-1)} & 2(n-2)& c\sqrt{3(d'-2)}&\ldots &0\\
            \vdots & \ddots & \ddots& \ddots & \ddots & \vdots \\
            0&0&\ldots & c\sqrt{2(d'-1)} &(d'-1)(n-2) & c\sqrt{d'}\\
            0 & 0 & \ldots & 0 & c\sqrt{d'} & d'(n-2)\\
          \end{array}\right)
\end{equation}
Where $c=\sqrt{n-1}$ and $d'$ is a parameter which is related to
the number of strata, for the first term in equation (3-48) in the
binary Hamming, the $d'$ is equal to $d$ and for the second term
in binary Hamming the parameter $d'=d-2$, therefore in each stage
this parameter decreases consecutively.

\subsubsection{Entanglement entropy in Hamming network between two equal parts: First half and second half strata}
In this section, we want to calculate bipartite entanglement
between two equal parts of strata in Hamming network. after
applying the generalized Schur complement method to the potential
matrices of each strata in Hamming networks, we have a $2\times 2$
matrix finally, which it's entries are
$$a_{12}=-2gc\sqrt{(\frac{d'+1}{2})(\frac{d'+1}{2})}=-gc(d'+1)=-g\sqrt{n-1}(d'+1)$$
$$a_{11}=x-\alpha_k-\frac{\omega_{k-1}}{x-\alpha_{k-1}-\frac{\omega_{k-2}}{\frac{\vdots}{x-\alpha_3-\frac{\omega_2}{x-\alpha_2-\frac{\omega_1}{x-\alpha_1}}}}}$$
Where $x=1+2gd(n-1)$,
$$\alpha_i=-2g(i-1)(n-2),\quad\quad i=1,2,\ldots,\frac{d'+1}{2}$$
\begin{equation}
\omega_i=4g^2c^2i(d'-i+1),\quad\quad i=1,2,\ldots,\frac{d'-1}{2}
\end{equation}
And
$$a_{22}=x-\alpha'_k-\frac{\omega'_{k-1}}{x-\alpha'_{k-1}-\frac{\omega'_{k-2}}{\frac{\vdots}{x-\alpha'_3-\frac{\omega'_2}{x-\alpha'_2-\frac{\omega'_1}{x-\alpha'_1}}}}}$$

$$\alpha'_i=-2g(d'-i+1)(n-2),\quad\quad i=1,2,\ldots,\frac{d'+1}{2}$$
\begin{equation}
\omega'_i=4g^2c^2i(d'-i+1),\quad\quad i=1,2,\ldots,\frac{d'-1}{2}
\end{equation}
Then the parameter $\gamma_i$ ($\gamma_i$ for $i$th stratum) for
each strata is obtained from (3-22):
\begin{equation}
\gamma=\frac{a_{12}}{\sqrt{a_{11}a_{22}}}
\end{equation}
\subsubsection{Entanglement entropy in Hamming network between two equal parts: even strata are in first subset and odd strata are in second subset}
The entanglement entropy and Schmidt number are considered between
two equal parts of hamming network, that even strata are in first
subset and odd strata are in second subset.In this kind of
partitioning we can consider for two case. First case for $d=odd$
and second case for $d=even$.
\\\textbf{Case I($d$ is odd):}In this case the potential matrix
$V_{11}$ can be written as following:
\begin{equation}
\left(%
\begin{array}{cccc}
  1+2gdc^{2} & 0 & ... & 0 \\
  0 & 1+2g(dc^{2}-2(c^{2}-1)) & ... & 0 \\
  \vdots & \vdots & \ddots & \vdots \\
  0 & ... & 0 & 1+2g(dc^{2}-(d-1)(c^{2}-1))\\
\end{array}%
\right)_{\frac{d+1}{2}\times\frac{d+1}{2}}
\end{equation}
And $V_{22}$ is
\begin{equation}
\left(%
\begin{array}{cccc}
  1+2g(dc^{2}-(c^{2}-1)) & 0 & ... & 0 \\
  0 & 1+2g(dc^{2}-3(c^{2}-1)) & ... & 0 \\
  \vdots & \vdots & \ddots & \vdots \\
  0 & ... & 0 & 1+2g(dc^{2}-d(c^{2}-1))\\
\end{array}%
\right)_{\frac{d+1}{2}\times\frac{d+1}{2}}
\end{equation}
 And the connection matrix of potential matrix $V_{12}$ is:
\begin{equation}
\left(%
\begin{array}{cccc}
  -2gc\sqrt{d} & 0 & ... & 0 \\
  -2gc\sqrt{2(d-1)} & -2gc\sqrt{3(d-2)}& ... & 0 \\
  \vdots & \vdots & \ddots & \vdots \\
  0 & ... & -2gc\sqrt{2(d-1)} & -2gc\sqrt{d} \\
\end{array}%
\right)_{\frac{d+1}{2}\times\frac{d+1}{2}}
\end{equation}
Therefore,we have $\frac{d+1}{2}$ parameters $\gamma$
$$\gamma_{1}=\frac{2g\sqrt{n-1}}{\sqrt{1+2g(n(d-1))}\sqrt{1+2g(n(d-1)-(n-2))}}$$
$$\gamma_{2}=\frac{(2+4)g\sqrt{n-1}}{\sqrt{1+2g(n(d-1)-2(n-2))}\sqrt{1+2g(n(d-1)-3(n-2))}}$$
$$\vdots$$
\begin{equation}
\gamma_{\frac{d+1}{2}}=\frac{(2+4\frac{d-1}{2})g\sqrt{n-1}}{\sqrt{1+2g(n(d-1)-(d-1)(n-2))}\sqrt{1+2g(n(d-1)-d(n-2))}}
\end{equation}
Where $c$ is $\sqrt{n-1}$.
\\\textbf{Case II($d$ is even):}In this
case the potential matrix $V_{11}$ is
\begin{equation}
\left(%
\begin{array}{cccc}
  1+2gdc^{2} & 0 & ... & 0 \\
  0 & 1+2g(dc^{2}-2(c^{2}-1)) & ... & 0 \\
  \vdots & \vdots & \ddots & \vdots \\
  0 & ... & 0 & 1+2g(dc^{2}-d(c^{2}-1))\\
\end{array}%
\right)_{\frac{d}{2}+1\times\frac{d}{2}+1}
\end{equation}
And $V_{22}$ is
\begin{equation}
\left(%
\begin{array}{cccc}
  1+2g(dc^{2}-(c^{2}-1)) & 0 & ... & 0 \\
  0 & 1+2g(dc^{2}-3(c^{2}-1)) & ... & 0 \\
  \vdots & \vdots & \ddots & \vdots \\
  0 & ... & 0 & 1+2g(dc^{2}-(d-1)(c^{2}-1))\\
\end{array}%
\right)_{\frac{d}{2}\times\frac{d}{2}}
\end{equation}
And the connection matrix of potential matrix $V_{12}$ is:
\begin{equation}
\left(%
\begin{array}{cccc}
  -2gc\sqrt{d} & 0 &... & 0 \\
  -2gc\sqrt{2(d-1)} & -2gc\sqrt{3(d-2)} & ... & 0 \\
  \vdots & \vdots & \ddots & \vdots \\
  0 & ... & -2gc\sqrt{2(d-1)} & -2gc\sqrt{d} \\
\end{array}%
\right)_{\frac{d}{2}+1\times\frac{d}{2}}
\end{equation}
Therefore, we have $\frac{d}{2}$ parameters $\gamma$
$$\gamma_{1}=\frac{4g\sqrt{n-1}}{\sqrt{1+2g(n(d-1))}\sqrt{1+2g(n(d-1)-(n-2))}}$$
$$\gamma_{2}=\frac{(4+4)g\sqrt{n-1}}{\sqrt{1+2g(n(d-1)-2(n-2))}\sqrt{1+2g(n(d-1)-3(n-2))}}$$
$$\vdots$$
\begin{equation}
\gamma_{\frac{d+1}{2}}=\frac{(2+4(\frac{d}{2}-1))g\sqrt{n-1}}{\sqrt{1+2g(n(d-1)-(d-2)(n-2))}\sqrt{1+2g(n(d-1)-(d-1)(n-2))}}
\end{equation}
 Finally the entropy of entanglement
of each strata, i.e. $S(\rho)$, is obtained from Eq.$(3-29)$. So
the total entropy is $S(\rho)=\Sigma_{i}S(\rho_i)$.
\subsubsection{Entanglement entropy in Hamming network between two equal parts of adjacency matrix}
In this part, we want to calculate the entanglement entropy in
Hamming network between two equal parts of adjacency matrix.
\\\textbf{Case I($n$ is even):} we can define the blocks of
adjacency matrix as following:
\begin{equation}
A_{11}=A_{22}=(J_{\frac{n}{2}}-I_{\frac{n}{2}})\otimes
I_{n}\otimes...\otimes I_{n}+I_{\frac{n}{2}}\otimes
(J_{n}-I_{n})\otimes...\otimes I_{n}+...+I_{\frac{n}{2}}\otimes
I_{n}\otimes...\otimes I_{n}\otimes(J_{n}-I_{n})
\end{equation}
And, the connection matrix is
\begin{equation}
A_{12}=J_{\frac{n}{2}}\otimes I_{n}\otimes...\otimes I_{n}
\end{equation}
So, we have the $d$ types of parameter $\gamma$
$$\gamma_{1}=\frac{ng}{1+ng}$$
$$\gamma_{2}=\frac{ng}{1+3ng}$$
$$\gamma_{3}=\frac{ng}{1+5ng}$$
$$\vdots$$
\begin{equation}
\gamma_{d}=\frac{ng}{1+(2d-1)ng}
\end{equation}
For $d=2$, the degeneracy number of $\gamma_{1}$ is $1$ and the
degeneracy number of $\gamma_{2}$ is $n^{d-1}-1$.
\\For $d\geq3$, the degeneracy numbers of $\gamma_{i}$ respectively are
\\$1$ , $(d-1)(n-1)$ , $(d-2)(n-1)^{2}$ , $(d-3)(n-1)^{3}$ , $...$
, $(d-(d-1))(n-1)^{d-1}$.
\\\textbf{Case II($n$ is odd):}we can define the blocks of
adjacency matrix as following:
\begin{equation}
A_{11}=(J_{\frac{n+1}{2}}-I_{\frac{n+1}{2}})\otimes
I_{n}\otimes...\otimes I_{n}+I_{\frac{n+1}{2}}\otimes
(J_{n}-I_{n})\otimes...\otimes I_{n}+...+I_{\frac{n+1}{2}}\otimes
I_{n}\otimes...\otimes I_{n}\otimes(J_{n}-I_{n})
\end{equation}
\begin{equation}
A_{22}=(J_{\frac{n-1}{2}}-I_{\frac{n-1}{2}})\otimes
I_{n}\otimes...\otimes I_{n}+I_{\frac{n-1}{2}}\otimes
(J_{n}-I_{n})\otimes...\otimes I_{n}+...+I_{\frac{n-1}{2}}\otimes
I_{n}\otimes...\otimes I_{n}\otimes(J_{n}-I_{n})
\end{equation}
And, the connection matrix is
\begin{equation}
A_{12}=J_{\frac{n+1}{2}\times\frac{n-1}{2}}\otimes
I_{n}\otimes...\otimes I_{n}
\end{equation}
Where $J_{\frac{n+1}{2}\times\frac{n-1}{2}}$ is the matrix with
$\frac{n+1}{2}$ rows and $\frac{n-1}{2}$ columns, with all
elements are $1$.
\\So, we have the $d$ types of parameter
$\gamma$
$$\gamma_{1}=\frac{\sqrt{(n+1)(n-1)}g}{\sqrt{1+(n+1)g}\sqrt{1+(n-1)g}}$$
$$\gamma_{2}=\frac{\sqrt{(n+1)(n-1)}g}{\sqrt{1+(3n+1)g}\sqrt{1+(3n-1)g}}$$
$$\gamma_{3}=\frac{\sqrt{(n+1)(n-1)}g}{\sqrt{1+(5n+1)g}\sqrt{1+(5n-1)g}}$$
$$\vdots$$
\begin{equation}
\gamma_{d}=\frac{\sqrt{(n+1)(n-1)}g}{\sqrt{1+((2d-1)n+1)g}\sqrt{1+((2d-1)n-1)g}}
\end{equation}
For $d=2$, the degeneracy number of $\gamma_{1}$ is $1$ and the
degeneracy number of $\gamma_{2}$ is $n^{d-1}-1$.
\\For $d\geq3$, the degeneracy numbers of $\gamma_{i}$ respectively are
\\$1$ , $(d-1)(n-1)$ , $(d-2)(n-1)^{2}$ , $(d-3)(n-1)^{3}$ , $...$
, $(d-(d-1))(n-1)^{d-1}$. Finally the entropy of entanglement of
each strata, i.e. $S(\rho)$, is obtained from Eq.$(3-29)$. So the
total entropy is $S(\rho)=\Sigma_{i}S(\rho_i)$.

\subsection{Entanglement entropy between all kinds of two parts
in Hamming H(2,3) network} In this section we separated Hamming
$H(2,3)$ network into to  parts,the first part contains $5$ nodes
and the second part contains $4$ nodes, then calculated the
entanglement entropy between all kinds of bisection numerically.
In this case, there are $126$ kinds of partitioning in a way that
two parts. Numerical calculations show that some of these $126$
sets have the same entropy, such that there are $5$ kinds of
different values for entanglement entropies.

\setlength{\unitlength}{0.75cm}
\begin{picture}(6,8)
\linethickness{0.075mm}

\put(2,2){\circle*{0.2}} \put(4,2){\circle*{0.2}}
\put(6,2){\circle*{0.2}} \put(2,4){\circle*{0.2}}
\put(4,4){\circle*{0.2}} \put(6,4){\circle*{0.2}}
\put(2,6){\circle*{0.2}} \put(4,6){\circle*{0.2}}
\put(6,6){\circle*{0.2}}

\put(1.7,6.1){$1$} \put(6.1,6.1){$2$} \put(4,6.1){$3$}
\put(2.1,3.5){$4$} \put(4.1,3.5){$7$} \put(6.1,3.6){$6$}
\put(2.1,2.1){$5$} \put(4.1,2.1){$9$} \put(6.1,1.6){$8$}

\put(2,2){\line(1,0){4}} \put(2,2){\line(0,1){4}}
\put(2,4){\line(1,0){4}} \put(4,2){\line(0,1){4}}
\put(2,6){\line(1,0){4}} \put(6,2){\line(0,1){4}}

\put(2,4){\oval(1,4)[l]} \put(4,4){\oval(1,4)[l]}
\put(6,4){\oval(1,4)[r]} \put(4,4){\oval(4,1)[t]}
\put(4,6){\oval(4,1)[t]} \put(4,2){\oval(4,1)[b]}

\put(0,0){\footnotesize  FIG I: Hamming $H(2,3)$ graph.}
\end{picture}

 The maximum entanglement entropy in this section is for subset:
$1$ of table $1$. In this case there are the maximum edges between
two parts. The minimum entanglement entropy in this section are
for subset:$5$ of table $1$. In these cases there are the minimum
number of edges between two parts.

\subsection{Entanglement entropy between all kinds of two parts
in Hamming H(2,4) network} In this section we separated Hamming
$H(2,4)$ network into to equal parts, then calculated the
entanglement entropy between all kinds of bisection numerically.
The order of name of vertices are based on
$A=I_{4}\bigotimes(J_{4}-I_{4})+(J_{4}-I_{4})\bigotimes I_{4}$.

In this case, there are $6435$ kinds of partitioning in a way that
two parts. Numerical calculations show that some of these $6435$
sets have the same entropy, such that there are $22$ kinds of
different values for entanglement entropies. The maximum entropy
is in the partition that the vertices $(1,2,5,7,10,12,15,16)$ in
the first part. And, the minimum entropy is in the partition that
the vertices $(1,2,3,4,5,6,7,8)$ in the first part and other
vertices is in the second part. In the table $2$, the assortment
of entropy from maximum to minimum is given respect to abundance
and agent of each group.
\begin{table}
\begin{tabular}{|c|c|}
\hline
set & partitions \\
\hline $1$   & $(2,3,4,5,6),(2,3,4,5,7),(2,3,4,5,8),(2,3,4,5,9),(2,3,4,6,9),(2,3,,4,7,8),(2,3,5,6,9),$  \\
               & $(2,3,5,7,8),(2,4,5,6,9),(2,4,5,7,8),(3,4,5,6,9),(3,4,5,7,8),(3,5,6,7,8),(2,4,7,8,9),$  \\
               & $(3,4,6,8,9),(2,5,6,7,9),(1,5,6,7,9),(1,5,6,7,8),(1,4,7,8,9),(1,4,6,8,9),(1,3,6,7,8),$  \\
               & $(1,3,6,8,9),(1,2,7,8,9),(1,2,6,7,9),(1,2,4,7,8),(1,3,5,7,8),(1,3,5,6,9),(1,3,5,6,7),$  \\
               & $(1,3,4,8,9),(1,3,4,6,9),(1,3,4,6,8),(1,2,5,6,7),(1,2,5,6,9),(1,2,5,7,9),(1,2,4,7,9),$  \\
               & $(1,2,4,8,9),$  \\

\hline       $2$   & $(2,3,4,8,9),(2,3,5,6,7),(2,4,5,7,9),(3,4,5,6,8),(1,6,7,8,9),(1,3,4,7,8),(1,3,5,6,9),$\\
                   & $(2,3,4,8,9),(1,2,4,6,9),$\\

\hline $3$   & $(3,4,5,8,9),(3,4,5,6,7),(2,4,5,8,9),(2,,4,5,6,7),(2,3,5,6,8),(2,3,5,7,9),(2,3,4,7,9),$\\
              & $(2,3,4,6,8),(2,4,6,7,9),(2,4,6,8,9),(2,,5,6,7,8),(2,5,7,8,9),(3,4,6,7,8),(3,4,7,8,9),$  \\
              & $(3,5,6,7,9),(3,5,6,8,9),(1,5,7,8,9),(1,5,6,8,9),(1,4,6,7,9),(1,4,6,7,8),(1,3,7,8,9),$  \\
              & $(1,3,6,7,9),(1,2,6,7,8),(1,2,6,8,9),(1,2,3,7,8),(1,4,5,6,9),(1,4,5,7,8),(1,2,3,6,9),$  \\
              & $(1,3,4,5,6),(1,3,4,5,8),(1,2,4,5,7),(1,2,4,5,9),(1,2,3,5,7),(1,2,3,5,6),(1,2,3,4,9),$  \\
              & $(1,2,3,4,8)$\\
\hline    $4$ & $(3,4,5,7,9),(2,4,5,6,8),(2,3,5,8,9),(2,3,4,6,7),(4,5,7,8,9),(4,5,6,8,9),(4,5,6,7,9),$ \\
              & $(4,5,6,7,8),(2,3,7,8,9),(2,3,6,8,9),(2,3,6,7,9),(2,3,6,7,8),(5,6,7,8,9),(4,6,7,8,9),$  \\
              & $(3,6,7,8,9),(2,6,7,8,9),(1,2,4,6,8),(1,4,5,6,8),(1,4,5,7,9),(1,3,5,7,9),(1,3,5,8,9),$  \\
              & $(1,3,4,7,9),(1,3,4,6,7),(1,2,5,6,8),(1,2,5,8,9),(1,2,3,6,7),(1,2,3,8,9),(1,2,4,6,7),$  \\
              & $(1,3,4,5,7),(1,3,4,5,9),(1,2,4,5,6),(1,2,4,5,8),(1,2,3,5,8),(1,2,3,5,9),(1,2,3,4,7),$  \\
              & $(1,2,3,4,6)$\\
\hline $5$   & $(2,4,6,7,8),(2,5,6,8,9),(3,4,6,7,9),(3,4,7,8,9),(1,4,5,8,9),(1,4,5,6,7),(1,2,3,6,8),$ \\
             & $(1,2,3,7,9),(1,2,3,4,5)$ \\
\hline
\end{tabular}
\caption{\label{entropy} All equal subsets in $H(2,3)$. The
entanglement entropy for these subsets are $S_1>S_2>S_3>S_4>S_5$.}
\end{table}

\begin{table}
\begin{tabular}{|c|c|c|}
\hline
set & abundance & agent of partitions \\
\hline $1$ & $36$   & $(1,2,6,7,10,12,15,16)$  \\

\hline   $2$ &     $9$   & $(1,2,5,6,11,12,15,16)$\\

\hline $3$ & $288$   & $(1,2,3,5,8,10,12,15)$\\
\hline  $4$ & $288$   & $(1,2,3,5,6,11,12,16)$\\
\hline  $5$ & $576$   & $(1,2,3,5,6,9,12,15)$\\
\hline  $6$ & $576$   & $(1,2,3,5,6,9,12,16)$\\
\hline $7$ & $432$   & $(1,2,3,5,6,8,11,16)$\\
\hline  $8$ & $288$   & $(1,2,3,5,6,9,11,16)$\\
\hline  $9$ & $24$   & $(1,2,3,5,6,7,12,16)$\\
\hline $10$ & $576$   & $(1,2,3,5,6,8,9,15)$\\
\hline $11$ & $288$   & $(1,2,3,5,6,7,9,16)$\\
\hline $12$ & $288$   & $(1,2,3,4,5,6,11,16)$\\
\hline $13$ & $72$   & $(1,2,3,5,6,8,9,14)$\\
\hline $14$ & $1152$   & $(1,2,3,4,5,6,9,15)$\\
\hline $15$ & $288$   & $(1,2,3,4,5,6,11,15)$\\
\hline $16$ & $72$   & $(1,2,3,4,5,6,11,12)$\\
\hline $17$ & $288$   & $(1,2,3,4,5,6,9,14)$\\
\hline $18$ & $288$   & $(1,2,3,4,5,6,9,11)$\\
\hline $19$ & $96$   & $(1,2,3,4,5,6,7,12)$\\
\hline $20$ & $360$   & $(1,2,3,4,5,6,7,9)$\\
\hline $21$ & $144$   & $(1,2,3,4,5,6,9,13)$\\
\hline $22$ & $6$   & $(1,2,3,4,5,6,7,8)$\\
\hline
\end{tabular}
\caption{\label{entropy} agent of equal subsets in $H(2,4)$}.
\end{table}
\section{Conclusion}
The entanglement entropy is obtained between two parts in the
Hamming networks that their nodes are considered as quantum
harmonic oscillators. The generalized Schur complement method is
used to calculate the Schmidt numbers and entanglement entropy
between two parts of Hamming graph. Analytically, entanglement
entropy in two special partitions are calculated and the maximum
entropy in different partitioning is given. Numerical results show
that, the maximum and minimum entanglement entropy have a relation
with the edges between two parts.
\\One expects that the entanglement entropy and Schmidt numbers
can be calculated in Johnson networks.
\section{Appendix}
\subsection{Other expression for stratification of Hamming graph}
In this section,we want obtain the matrix $A_{ij}$ by straight
way.We know the basis of different strata is
\begin{equation}
{|0 0 ... 0\rangle},{|1 0 ... 0\rangle,|0 1 ... 0\rangle,...,|0 0
... 1\rangle},...,{|1 1 ... 1\rangle}\end{equation}So,we know
basis of each strata from number of $1$.in $m$'th strata,the basis
are:$e_{i_{1},i_{2},...,i_{m}}$, where
$i_{1}<i_{2}<...<i_{m}$.Therefore, the adjacency matrix is defined
as following
\begin{equation}
A_{i_{1},i_{2},...,i_{m},j_{1},j_{2},...,j_{m+1}}=\delta_{i_{1}j_{1}}\delta_{i_{2}j_{2}}...\delta_{i_{m}j_{m}}\prod_{k=1}^{m}(1-\delta_{i_{k}j_{m+1}})+
\delta_{i_{1}j_{2}}\delta_{i_{2}j_{3}}...\delta_{i_{m}j_{m+1}}\prod_{k=1}^{m}(1-\delta_{i_{k}j_{k}})+...\end{equation}
From the equation $(3-32)$, The adjacency matrix in Hamming graph
commutate with permutation.If $\pi$ is permutation, we have
\begin{equation}
[A,\pi]=0
\end{equation}
It means that,the adjacency matrix under displacement of particles
is invariant.
\begin{equation}
\pi_{m}A=A\pi_{m+1}
\end{equation}
Let $|\psi\rangle$ is the state, that is invariant under
permutation of $\pi_{m+1}$
\begin{equation}
\pi_{m+1}|\psi\rangle=\theta|\psi\rangle
\end{equation}
That $\theta$ is the phase.So, the acting of $A$ on the
$|\psi\rangle$ is the eigenvector of $\pi_{m}$.so, that is enough
finding the eigenvectors of permutation group.And, we know that
the eigenvectors of permutation group is discrete fourier
transform.So,we have
\begin{equation}
|i_{1},i_{2},...,i_{m}\rangle=\frac{1}{\sqrt{m!d^{m}}}\sum_{\alpha_{1},\alpha_{2},...,\alpha_{m}}\varepsilon_{\alpha_{1},\alpha_{2},...,\alpha_{m}}\omega^{(i_{\alpha_{1}}-1)(j_{1}-1)+(i_{\alpha_{2}}-1)(j_{2}-1)+...+(i_{\alpha_{m}}-1)(j_{m}-1)}|0\overbrace{1}^{i_{1}}0...\overbrace{1}^{i_{2}}0...\rangle\end{equation}
Where $\varepsilon$ is Levi-Civita symbol

\subsection{Finding the normalization factor for strata of Hamming network}
Let
\begin{equation}|\widetilde{\phi_{2}}\rangle_{k}=\sum_{i_{1}<i_{2}}(\omega_{d}^{ki_{1}}+\omega_{d}^{ki_{2}})|0...0\overbrace{1}^{i_{1}}0...0\overbrace{1}^{i_{2}}0...0\rangle\end{equation}
Now,we must find the normalization multiplier of
$|\widetilde{\phi_{2}}\rangle_{k}$.Therefore, we
define$|\widetilde{\phi_{2}}\rangle_{k'}$ orthogonal to
$|\widetilde{\phi_{2}}\rangle_{k}$.
\begin{equation}|\widetilde{\phi_{2}}\rangle_{k'}=\sum_{i_{1}<i_{2}}(\overline{\omega}_{d}^{k'i_{1}}+\overline{\omega}_{d}^{k'i_{2}})|0...0\overbrace{1}^{i_{1}}0...0\overbrace{1}^{i_{2}}0...0\rangle\end{equation}
Therefore
$$_{k}\langle\widetilde{\phi_{2}}|\widetilde{\phi_{2}}\rangle_{k'}=\frac{1}{2}\sum_{i_{1}\neq
i_{2}}(\omega_{d}^{ki_{1}}+\omega_{d}^{ki_{2}})(\overline{\omega}_{d}^{k'i_{1}}+\overline{\omega}_{d}^{k'i_{2}})=$$
\begin{equation}\frac{1}{2}((d-2)\omega^{(k-k')i_{1}}+(d-2)\omega^{(k-k')i_{2}})=d(d-2)\end{equation}
So, the $|\widetilde{\phi_{2}}\rangle_{k}$ is
\begin{equation}
|\widetilde{\phi_{2}}\rangle_{k}=\frac{1}{\sqrt{d(d-2)}}\sum_{i_{1}<i_{2}}(\omega_{d}^{ki_{1}}+\omega_{d}^{ki_{2}})|0...0\overbrace{1}^{i_{1}}0...0\overbrace{1}^{i_{2}}0...0\rangle\end{equation}
finally, by the same way,we have
\begin{equation}
|\widetilde{\phi_{m}}\rangle_{k}=\sum_{i_{1}<i_{2}<...<i_{m}}(\omega_{d}^{ki_{1}}+\omega_{d}^{ki_{2}}+...+\omega^{ki_{m}})|0...0\overbrace{1}^{i_{1}}0...0\overbrace{1}^{i_{2}}0...0\overbrace{1}^{i_{m}}0...0\rangle\end{equation}
and
\begin{equation}
|\widetilde{\phi_{m}}\rangle_{k'}=\sum_{i_{1}<i_{2}<...<i_{m}}(\overline{\omega}_{d}^{k'i_{1}}+\overline{\omega}_{d}^{k'i_{2}}+...+\overline{\omega}^{k'i_{m}})|0...0\overbrace{1}^{i_{1}}0...0\overbrace{1}^{i_{2}}0...0\overbrace{1}^{i_{m}}0...0\rangle\end{equation}
Therefore
\begin{equation}_{k}\langle\widetilde{\phi_{m}}|\widetilde{\phi_{m}}\rangle_{k'}=\frac{1}{m!}\sum_{i_{1}\neq
i_{2}\neq ... \neq
i_{m}}(\omega_{d}^{ki_{1}}(\overline{\omega}_{d}^{k'i_{1}}+\overline{\omega}_{d}^{k'i_{2}}+...+\overline{\omega}_{d}^{k'i_{m}})=
\frac{d(d-2)(d-3)...(d-m-1)(d-m)}{(m-1)!}\end{equation} So, the
$|\widetilde{\phi_{m}}\rangle_{k}$ is
\begin{equation}
|\widetilde{\phi_{m}}\rangle_{k}=\sqrt{\frac{(m-1)!}{d(d-2)...(d-m)}}\sum_{i_{1}<i_{2}<...<i_{m}}(\omega_{d}^{ki_{1}}+\omega_{d}^{ki_{2}}+...+\omega^{ki_{m}})|0...0\overbrace{1}^{i_{1}}0...0\overbrace{1}^{i_{2}}0...0\overbrace{1}^{i_{m}}0...0\rangle\end{equation}


\begin{thebibliography}{99}
\bibitem{1} M. A. Nielsen and I. L. Chuang, Quantum Computation and Quantum Information,
(Cambridge Univ. Press, 2000).
\bibitem{2} A. Peres, Quantum Theory: Concepts and Methods, (Kluwer Academic Publishers, Dordrecht, 1993).
\bibitem{3} J. Preskill, Quantum Computation, (1997).
\bibitem{4} D. Bouwmeester, A. Ekert, and A. Zeilinger (Eds.), The Physics of Quantum Information,
(Springer, Berlin, 2000).
\bibitem{5} G. Alber, R. Beth, M. Horodecki, P. Horodecki, R. Horodecki, M. Rotteler, H. Weinfurter,
R. Werner, and A. Zeilinger (Eds.), Quantum Information,
(Springer-Verlag, Berlin, 2001).
\bibitem{6} L. E. Ballentine, Am. J. Phys. 55, 785 (1986).
\bibitem{7} R. Horodecki, P. Horodecki, M. Horodecki, K. Horodecki, Rev. Mod. Phys. 81:865-942, (2009).
\bibitem{8} M. B. Plenio, S. Virmani, Quantum Inf. Comput. 7, 1 (2007).
\bibitem{9} M. J. Donald, M. Horodecki, and O. Rudolph, J. Math. Phys. 43, 4252–4272 (2002).
\bibitem{10} J. Sperling, W. Vogel, Phys. Scr. 83, 045002 (2011).
\bibitem{11} Y. Guo and H. Fan, Quant-ph: 1304.1950 (2013).
\bibitem{12} J. M. Matera, R. Rossignoli, and N. Canosa, Phys. Rev. A 86, 062324 (2012).
\bibitem{13} G. Adesso, S. Ragy, A. R. Lee, Open Syst. Inf. Dyn. 21, 1440001 (2014).
\bibitem{114} O. Cernotík and J. Fiurášek, Phys. Rev. A 89, 042331 (2014).
\bibitem{15} G. Adesso and S. Piano, Phys. Rev. Lett 112, 010401 (2014).
\bibitem{16} F. Nicacio and M. C. de Oliveira, Phys. Rev. A 89, 012336 (2014).
\bibitem{17} D. Buono, G. Nocerino, S. Solimeno and A. Porzio, Laser Phys. 24 074008 (2014).
\bibitem{18} A. Cardillo, F. Galve, D. Zueco, J. G. Gardenes, Phys. Rev. A 87, 052312 (2013).
\bibitem{19} M.A.Jafarizadeh, S.Nami, F.Eghbali, Quant-ph:
1407.4044(2014).
\end{thebibliography}
\end{document}